\documentstyle[12pt,aaspp4,epsf]{article}

\begin{document}

\title{The Effects of Clumping and Substructure on ICM Mass Measurements}

\author{B. Mathiesen, A. E. Evrard}
\affil{Dept. of Physics, University of Michigan, Ann Arbor, MI 48109}
\author{J. J. Mohr\footnote{Chandra Fellow}}
\affil{Dept. of Astronomy and Astrophysics, University of Chicago, Chicago, IL 60637}

\begin{abstract}
We examine an ensemble of 48 simulated clusters to determine
the effects of small-scale density fluctuations and large-scale
substructure on X-ray measurements of the intracluster medium (ICM) mass.
We measure RMS density fluctuations in the ICM which can be characterized
by a mean mass-weighted clumping factor
$C \equiv \langle\rho^2\rangle/\langle\rho\rangle^2$ between 1.3 and 1.4
within a density contrast of 500.  These fluctuations arise from the
cluster history of accretion shocks and major mergers, and their presence
enhances the cluster's luminosity relative to the smooth case. We expect,
therefore, that ICM mass measurements utilizing models which assume uniform
density at a given radius carry a bias of order $\sqrt{C} \approx 1.16$.
We verify this result by performing ICM mass measurements on X--ray images
of the simulations and finding the expected level of bias.

The varied cluster morphologies in our ensemble also allow us to investigate
the effects of departures from spherical symmetry on our measurements.
We find that the presence of large-scale substructure does not further bias
the resulting gas mass unless it is pronounced enough to produce a second
peak in the image of at least 1\% the maximum surface brightness. We
analyze the subset of images with no secondary peaks and find a bias of
9\% and a Gaussian random error of 4\% in the derived mass.
\end{abstract}


\section{Introduction}

Studies of the intracluster medium (ICM) address many important cosmological
and astrophysical questions. Precision measurements of individual ICM
spectra and metallicities can be used to constrain models of massive star
formation. The slope and evolution of cluster scaling relations (e.g. the
luminosity vs. temperature or ICM mass vs. temperature) can provide insight
into the nature and importance of galactic winds, as well as the background
cosmology in which the ICM develops. In addition, current models of
structure formation suggest that the mass components of clusters are
representative of the universe as a whole; for this reason there has been
much effort expended towards measuring the contributions of dark matter,
ICM, and galaxies in the population.  The mean ICM mass fraction is
therefore a useful lower limit on the global baryon fraction, and can
also be used to constrain $\Omega_0$ if combined with primordial
nucleosynthesis calculations.

Many studies have been carried out recently which attempt to measure
the form of the ICM density profile using X--ray images (e.g. Mohr, Mathiesen,
\& Evrard 1999 (MME); White, Jones, \& Forman 1997;  Loewenstein \& 
Mushotzky 1996; David, Jones, \& Forman 1995, White \& Fabian 1995).
Following the lead of early work in the field (Forman \& Jones 1982), such
studies typically create an analytic model for the azimuthally averaged ICM
density profile, and use the X--ray surface brightness to constrain its
parameters.  These models generally assume that the ICM is uniform at a
given radius, that the density profile decreases monotonically from the
center, and that the cluster is spherically symmetric.

These assumptions break down in real clusters; even apparently relaxed
clusters tend to exhibit small asphericities. The prevalence of accretion
events and major mergers in the local population is also well established,
which calls into question the validity of a monotonic density profile. The
assumption of a uniform ICM is a reasonable approximation, but it too must
fail on sufficiently small scales. A certain level of random density
fluctuations is to be expected, and even minor accretion events are found in
simulations to produce persistent mild shocks and acoustic disturbances in
the ICM. The small scatter in observed scaling relations, for example
the ICM mass and mass fraction vs. temperature (MME), implies that these
assumptions produce results with at least 20\% precision; this fact inspires
confidence. It is important, however, to directly measure the systematic
errors incurred by these inaccuracies.

In accordance with the increasing quality of X--ray observations, the
scientific community has recently begun to address these issues. A number
of groups are investigating the possibility of a multiphase medium
(Thomas 1998, Lima Neto et al. 1997, Waxman \& Miralde-Escude 1995),
although most of the work to date focuses on cooling flows rather than
global ICM properties. A notable exception is the work of Gunn \& Thomas
(1996), who examine the effect of a multiphase ICM on baryon fraction
measurements. There have also been attempts to get around the assumption of
spherical symmetry by using the isophotal area (Mohr \& Evrard 1997) or
elliptical isophotes (Knopp \& Henry 1996, Buote \& Canizares 1996) to
constrain a density model. All of these methods, however, still fall prey to
one or more of the assumptions mentioned above.

Hydrodynamic simulations of cluster evolution are ideally suited to the
task of estimating the systematic errors incurred by these assumptions. 
In this study, we produce realistic X--ray surface brightness images from
a set of simulations and measure ICM masses by fitting the emission profile
to a spherically symmetric beta model for the density. Our independent
knowledge of the three-dimensional structure allows us to observe the effects
of gross substructure and small-scale density fluctuations on the derived
ICM mass directly, and correlate these errors with the kind of substructure
present. In \S 2 of this paper, we present some detail on the nature of the
simulations and how we create our X--ray images. In \S 3, we describe
how we analyze these images and extract ICM density profiles.  In
\S 4, we compare the ``observed'' masses to the simulations and
correlate the errors with structural properties of the ICM. Finally, in
\S 5 we restate our results and remark on how they can be applied to real
observations. Our results are phrased throughout in a manner indpendent of
the Hubble constant $H_0$.

\section{Data}

We use an ensemble of 48 hydrodynamical cluster simulations, divided
among four different cold dark matter (CDM) cosmological models.
These models are (i) SCDM ($\Omega_0 = 1$, $\sigma_8 = 0.6$, $h
= 0.5$, $\Gamma = 0.5$); (ii) $\tau$CDM ($\Omega_0 = 1$, $\sigma_8 = 0.6$,
$h = 0.5$, $\Gamma = 0.24$); (iii) OCDM ($\Omega_0 = 0.3$,
$\sigma_8 = 1.0$, $h = 0.8$, $\Gamma = 0.24$); and (iv) 
$\Lambda$CDM ($\Omega_0 = 0.3$,  $\lambda_0 = 0.7$, $\sigma_8 = 1.0$,
$h = 0.8$, $\Gamma = 0.24$). Here the Hubble constant is
$100h$ km s$^{-1}$ Mpc$^{-1}$, and $\sigma_8$ is the power spectrum
normalization on $8h^{-1}$ Mpc scales. The initial conditions are Gaussian
random fields consistent with a CDM transfer function with the specified
$\Gamma$ (Davis et al. 1985). The baryon density is set in each case to a 
fixed fraction of the total density ($\Omega_b = 0.2\Omega_0$). The
simulation scheme is P3MSPH; first a P$^3$M (DM only) simulation is used
to find cluster formation sites in a large volume, then a hydrodynamic
simulation is performed on individual clusters to resolve their DM halo
and ICM structure in detail. The resulting cluster sample covers a little
more than a decade in total mass, ranging from about $10^{14}$ to
$2 \times 10^{15} M_{\odot}$.

We work with X--ray surface brightness maps derived from the simulations
according to procedures described by Evrard (1990a,b). The particles
representing pieces of the ICM are treated as bremsstrahlung emitters
at the local temperature and density, and this emission is distributed
over the image pixels according to a two-dimensional Gaussian with a
width equal to that of the particle's SPH smoothing kernel. Emission is
collected in the energy band [0.1,2.4] keV, and the resulting surface
brightness in each pixel is converted to counts per second according
to an approximate ROSAT PSPC energy conversion factor. We then ``observe''
this model image using an effective area map of the PSPC and an exposure
sufficient to yield $10^4$ cluster photons, comparable to ROSAT exposures
of clusters in the Edge sample (Edge et al. 1990). Finally, each pixel
is given a Poisson uncertainty based on the number of photons and the whole
image is smoothed on a scale of  $24.4\arcsec$ to improve the
signal-to-noise ratio. Angular distances are calculated as $d_A = cz/H_0$
with $z = 0.06$, so the physical smoothing scale is either 42.6 or 26.6 kpc.
Note that beause we are trying to isolate the systematic errors due to
substructure, we do not add Poisson noise to the image; the error bars are
only used to assign appropriate weights to the data in the analysis.

It should be noted that the simulations do not include certain processes known
to be present in real clusters, such as radiative cooling and the injection
of gas and energy by galactic supernovae. The cooling time for at least 
99\% of the SPH particles is much longer than the Hubble time, and would
have little effect on the structure of the ICM; nevertheless, our simulations
cannot develop cooling flows.  Detailed studies of real cooling flows
find mass deposition rates no larger than hundreds of solar masses per
year,  suggesting that they comprise only a small fraction of the ICM
mass (White et al. 1998). Simulations which include galactic winds
create clusters with more realistic density profiles, but the winds
are not found to greatly affect the temperature structure of the gas
(Metzler \& Evrard 1994). It is conceivable that the presence of galactic
winds would also create more clumping in the ICM, but this effect is
probably negligible next to the variations caused by accretion events.

The clusters display great morphological diversity, ranging from examples
which appear almost perfectly spherical and relaxed to others which are
undergoing a three-way merger event. Each cluster is imaged in three
orthogonal projections, allowing us to compare the biases caused by
substructure along the line-of-sight and substructure in the plane of the
sky. We use all three projections in general, but confine ourselves to
one projection per cluster when making statistical comparisons to insure
that the probabilities are derived from statistically independent data. 
Further details on the nature of these simulations can be found in both
MME and Mohr \& Evrard (1997).

\section{Image Analysis}

We fit the emission to a beta model, with three-dimensional density
profile $\rho(r) = \rho_0[1+(r/r_c)^2]^{-3\beta/2}$ (Cavaliere \&
Fusco-Femiano, 1978). In this model $r_c$ and $\beta$ are free parameters
to be constrained by the X--ray emission profile, and $\rho_0$ is found by
normalizing the emission integral to the bolometric luminosity of the cluster.

Producing a cluster emission profile appropriate for fitting
requires three steps. First, the emission center is found by sliding a
circular aperture of radius 10 pixels over the image and minimizing the
distance between the center of the aperture and the centroid of the X-ray
photon distribution within the aperture. This approach converges even in
cases where the cluster emission is significantly skew (Mohr, Fabricant,
\& Geller 1993). Second, we calculate the azimuthally averaged radial
profile. Finally, each point in the profile is assigned a Poisson uncertainty
based on the number of photons in the annulus; this allows the fitting program
to assign approprate relative weights to the datapoints.

The PSPC point spread function (PSF) reduces a cluster's central
intensity and increases the apparent value of $r_c$.  This effect
changes the best-fit values of our model parameters and biases
the derived ICM mass if left untreated. The incurred error is potentially
significant for clusters with small angular diameter or pronounced
cooling flows, and is simulated here by the image smoothing mentioned
in \S 2.  Thus, rather than fitting the beta model directly to our profile,
we first convolve it with a PSF appropriate for our smoothing kernel.
Details on the mathematics of this technique can be found in MME and
Saglia et al. (1993). 

The fitting was performed from the center of the cluster out to $r_{500}$,
the radius at which the mean interior density is 500 times the critical
density. This measure was chosen to probe the easily visible region of each
cluster given typical backgrounds and sensitivities, as well as for physical
reasons which will become apparent in the next section.  By fitting out to
a fixed fraction of the virial radius rather than a fixed metric radius, we
insure that we are probing regions with similar dynamical gravitational
time scales. The ratio $\Upsilon$ between cluster baryon fractions and
the universal baryon fraction has also been calibrated by simulations;
a bias factor of $\Upsilon = 0.9 \pm 0.1$ was found at $r_{500}$
(Evrard 1997). Recent work with an independent simulation has
corroborated this result, finding $\Upsilon = 0.92 \pm 0.06$ at $r_{200}$
(Frenk et al. 1999). The radius $r_{500}$ is found in our simulations to
scale as $1.4 h^{-1} (T/10\mathrm{keV})^{1/2}$ Mpc, regardless of cosmology.

Finally, we checked our process by fitting 100 monte carlo images of true
beta models with similar smoothing and individual realizations of Poisson
noise, and found a chi-squared distribution consistent with a perfect
match between the fitting function and the underlying model. This
confirms that the high chi-squared values obtained in some images
result from real physical deviations from a spherical beta model. A more
complete discussion of our method, as applied to real PSPC data, is given
in MME.

\section{Results}

We measure the variance of density fluctuations in spherical shells
around the ICM particle with the lowest gravitational potential. This
variance is expressed in terms of a clumping factor $C$,
\begin{equation}
C \equiv \langle\rho^2\rangle/\langle\rho\rangle^2. 
\end{equation}
Each ICM particle in the simulation has mass $m_{\mathrm{gas}}$.
For a given shell lying between radii $r_1$ and $r_2$ with $N_{12}$ gas
particles, $\langle\rho\rangle$ is defined as the total mass in gas
particles divided by the volume of the shell, $m_{\mathrm{gas}}N_{12}/V$.
The numerator is calculated as
\begin{equation}
\langle\rho^2\rangle \equiv \frac{1}{V_{12}}\int^{r_2}_{r_1}d^3r\rho^2
\rightarrow\frac{m_{\rm gas}}{V_{12}}\sum_i^{N_{12}}\rho_i,
\end{equation}
with the right-hand term being the Lagrangian limit of the integral
and $\rho_i$ the SPH gas density of particle $i$ (Evrard 1988).

The results of this analysis are displayed in Figure~1. Although more
complicated structure is visible in some of our clusters, the average
magnitude of density fluctuations interior to $r_{500}$ seems to
be a slowly increasing function of radius. The radial bins are chosen
such that they represent fixed logarithmic intervals in the density
constrast $\delta_c$. Dramatic increases in clumping are occasionally
seen between $r_{500}$ and the more traditional virial radius; this
is another reason that for choosing to work within a density constrast
of 500.  Extending our analysis to the virial radius would return
essentially the same results with lower statistical significance.

The presence of these fluctuations enhances the cluster luminosity over
what would be expected for a smooth, single-phase ICM. The emissivity
in a given shell is given by
\begin{equation}
\varepsilon(r) = r^2dr\int d\Omega \frac{\rho^2(r,\Omega)}{\mu_e\mu_Hm_p^2}
\Lambda(T,\Omega).
\end{equation}
If the density fluctuations are modest, the temperature and ionization
structure of the gas in a given shell will be approximately uniform.
Taking these factors outside the integral and using equation (1) to
incorporate our clumping factor, we have
\begin{equation}
\varepsilon(r) = 4\pi r^2dr\frac{\Lambda(T)}{\mu_e\mu_Hm_p^2}
\langle\rho(r)\rangle^2 C(r).
\end{equation}
$C$, therefore, represents the factor by which a shell's emissivity is
enhanced over the single-phase case. This formalism can be invoked to
describe a true multiphase medium under arbitrary constraints
by choosing an appropriate function $C(r)$. Nagai, Evrard, \& Sulkanen
(1999) look at such distributions in more detail,
investigating the observable consequences of applying isobaric
multiphase models.

For each cluster we calculate the mass-weighted mean $\bar{C}$ over all
shells as an approximation to the overall bias on its total luminosity
and inferred core density $\rho_0^2$. These values, averaged again over
all clusters within a cosmological model, are as follows: 
$\bar{C}_o = 1.40$, $\bar{C}_s = 1.32$, $\bar{C}_{\Lambda} = 1.29$, and
$\bar{C}_{\tau} = 1.38$. The mean values of $M_{\beta}/M_{\mathrm{true}}$
for the four cosmologies are 1.34 (OCDM), 1.13 (SCDM), 1.11 ($\Lambda$CDM),
and  1.14 ($\tau$CDM). The mass error distributions for the four cosmologies
are mutually consistent, so we combine them to determine a mean ICM bias
for the entire ensemble of $1.18 \pm 0.02$.  We also calculate the ensemble
mean of $\bar{C}$, arriving at $1.34 \pm 0.02$. Since the total luminosity
is proportional to the square of the central gas density, we expect an
overall bias in our ICM mass estimates of $\bar{C}^{1/2}$, or $1.16 \pm
0.01$. The error bars given here are one standard deviation of the mean,
so this is excellent agreement.  We do not, however, find a strong
correspondence between the level of clumping and the ICM mass error in
individual clusters, however. Variations in the shape of the mass profile
and large-scale asymmetries, which contribute a random component to the
error, mask this relationship.  

It is difficult to judge what weighting scheme is most appropriate for
this calculation, since the pivotal regions of the emission profile
vary from cluster to cluster. The central points are very accurate and
do much to constrain the fit, but are few in number; the outer regions
have large error bars but many more data points.  We began our anaysis
with a mass-weighted clumping factor under the simple reasoning that it 
would greatly favor neither region. We also performed the exercise
of weighting the shells by their luminosity (which gives more importance
to the core) and volume (which emphasizes the outskirts). The resulting
ensemble mean clumping factors do not vary greatly under the different
weighting schemes: we found averages of $\bar{C} = 1.27$ for
luminosity-weighted shells and $\bar{C} = 1.39$ for volume-weighted shells.

We also attempt to identify correlations of mass error with large-scale
substructure signatures.  A surprising result is that we find no correlation
of $M_{\beta}/M_{\rm true}$ with centroid shift, which has been found
to be a good indicator of a cluster's dynamical state (Mohr, Evrard,
Fabricant, \& Geller 1995). Even those clusters with the highest centroid
shifts don't show a significantly different gas mass bias from the rest of
the sample. Apparently, azimuthally averaging the cluster profile compensates
for mild asphericities very well. We also looked for correlations of
mass error with various bulk properties of the ICM such as mass, 
emission-weighted temperature, and luminosity, finding none.
There is evidence for a weak correlation of mass error with the
model parameters $\beta$ and $r_c$, but this can be entirely attributed
to the tendency for strongly bimodal clusters to have profiles which
are not well fit by the model. Although our clusters typically
have steeper profiles than are observed in reality (c.f.), the lack
of shape dependence indicates that this is probably not a problem.
In none of the above tests were there discernible differences among
the four cosmologies.

In fact, we found only one ICM property to be correlated with $\delta M$.
Defining a ``secondary peak'' as a local maximum in the surface brightness
at least 1\% of the global maximum, we find that the most important
property is the existence of secondary peaks within $r_{500}$. We attempted
to further quantify the degree of asymmetry caused by subclumping, but
the mass error turned out to be uncorrelated with the number and strength
of the subpeaks. We therefore divide our ensemble into two subsets on this
basis, hereafter referred to as ``regular'' and ``bimodal'' clusters for
ease of language. Note that the subset of ``regular'' clusters contains
examples with large centroid shifts or asymmetries, and the ``bimodal''
subset includes a few trimodal clusters as well. The two subsets comprise 62\%
and 38\% of the total, respectively. Figure~2 displays a histogram
of ICM mass errors for all the images in our ensemble. The shaded region
corresponds to the regular clusters, while the remaining subset represents
the bimodal sample.

The bimodal population has a distribution of mass errors with mean
$\delta M = 0.188$ ($\delta M = 1 - M_{\beta}/M_{\rm true}$) and
standard deviation $\sigma = 0.084$. The remaining population has a
distribution with $\bar{\delta M} = 0.093$ and $\sigma = 0.041$. Applying
the K--S test, we find that the two populations are drawn from different
distributions with very high confidence (greater than 5$\sigma$) and that the
subset of non-bimodal images has an error distribution consistent with
the Gaussian form. In applying the K--S tests we worked with a subset of the
data consisting of just one image from each simulation, in order to maintain
statistical independence. The means and standard deviations just quoted,
however, reflect the distribution of the entire image ensemble. The smaller
samples in each category have distributions entirely consistent with the
complete sample, but we feel that the stated values better reflect the
size of real uncertainties by taking advantage of multiple projections.

Within our sample there are also 17 (out of 144) images which
contain a major subclump in line-of-sight with the primary cluster
and indistinguishable as a secondary peak. (A major subclump is one
which produces a secondary peak when the cluster is viewed from a
different angle.) These images, which occur in both the regular and
bimodal populations, have a distribution in $\delta M$ consistent
with the remaining set of bimodal images. This is reasonable given
the similarity of their three-dimensional physical structures. Their
partial membership in the set of regular images, however, does not
change the shape or mean of that distribution significantly. Such
occurences seem to be sufficiently uncommon that our heuristic
classification system remains useful.  Since strongly bimodal clusters
are more uncommon than our simulations indicate, real samples will
probably suffer even less from this contamination.  

\section{Conclusions}

Assuming a density profile which is uniform at a given radius
introduces a significant bias into measurements of the ICM mass, because
it ignores the presence of luminosity enhancements from overdense regions.
Our simulations indicate that these small-scale fluctuations produce a mean
overestimate of $\sim 10\%$ when we confine our analysis to regular
clusters. The advent of spatially resolved  X--ray spectral
imaging should allow us to test for irregularity in ICM structures
and constrain the level of these fluctuations with direct observations;
in the meantime it seems prudent to apply a correction of this scale to
current measurements, as is done in MME.  The relationship between these
fluctuations and the cluster accretion history remains an open question,
but this analysis suggests that there is no strong connection.

Applying our spherically symmetric model to a cluster's azimuthally
averaged surface brightness profile returns reasonably accurate ICM mass
measurements even in clusters exhibiting significant asymmetries, so long
as there are no significant secondary peaks in the image. The assumption
of spherical symmetry, while formally invalid in most cases, seems to
be borne out in practice in that the process of azimuthal averaging
returns a mean density profile which is unbiased by the presence of
moderate substructure. Such deviations from spherical symmetry contribute
a random error of $\sim 5\%$ to ICM mass measurments under our method.
The subset of bimodal (or multi-peaked) clusters has a much larger bias
and dispersion, and is perhaps best excluded from population studies
of the ICM.

\acknowledgments

This research was supported by NASA grants NAG5-2790, NAG5-3401,
and NAG5-7108, as well as NSF grant AST-9803199.  JJM is supported
through Chandra Fellowship grant PF8-1003, awarded through the Chandra
Science Center.  The Chandra Science Center is operated by the
Smithsonian Astrophysical Observatory for NASA under contract NAS8-39073.

\newpage

\figcaption[C.allmodels.ps]{The rms density fluctuations in spherical
shells for all 48 simulated clusters. The magnitude of these fluctuations
does not appear to depend significantly on the cosmological background
and is reasonably stable within $r_{500}$.  The dotted line represents the
mass-weighted mean $\bar{C}$, averaged over all clusters in that cosmology.}

\figcaption[mfracs_fig.ps]{The distribution of systematic errors in
ICM mass measurements based on the beta model. The full histogram
is for the entire image ensemble (three per cluster); the shaded
region is for the subset of regular cluster images.  The curve is
a Gaussian distribution with the same mean and standard deviation as the
shaded histogram. There is an outlier at $M_{\beta}/M_{\rm true}=1.8$
which lies beyond the limits of the plot.}

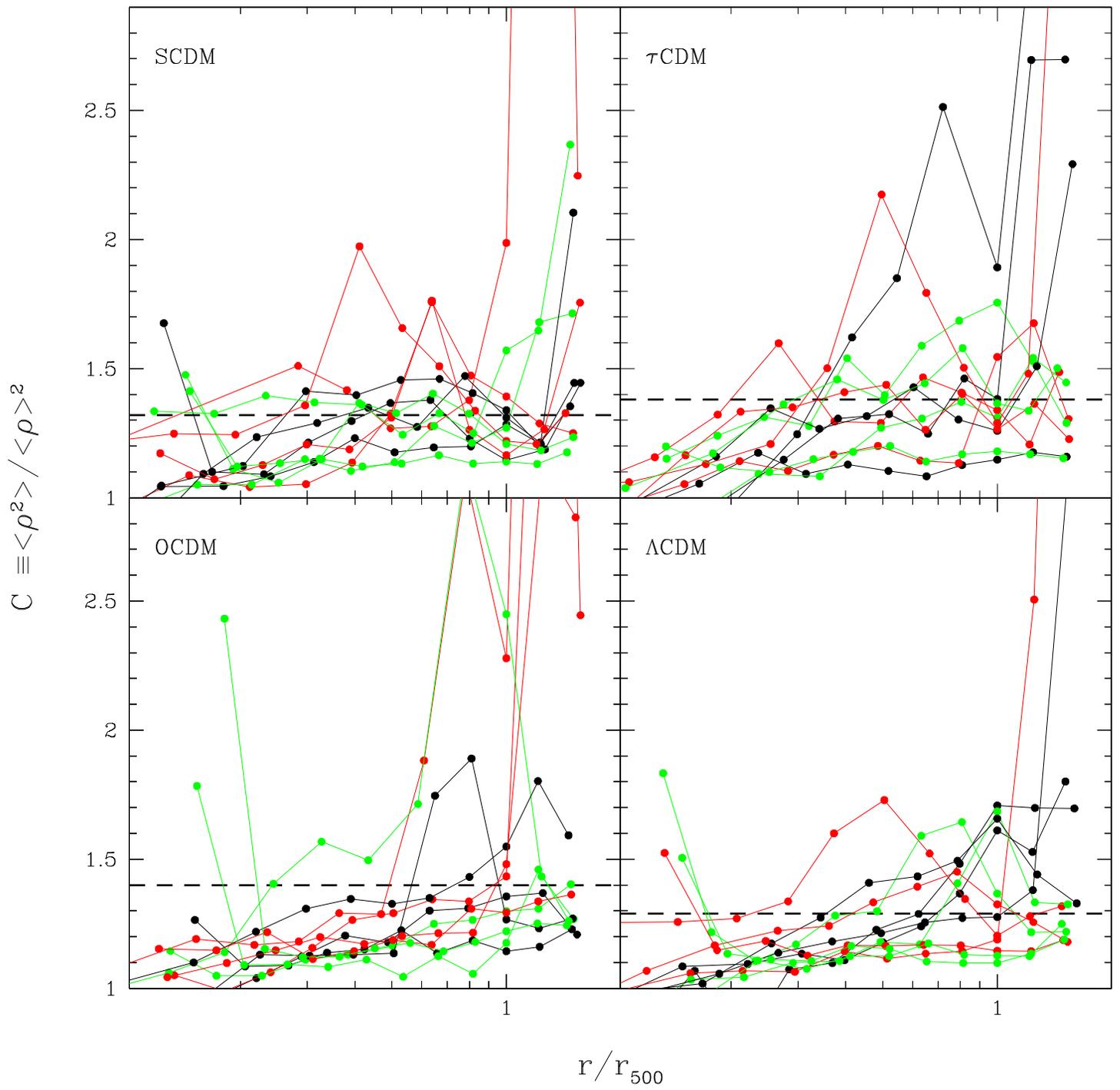
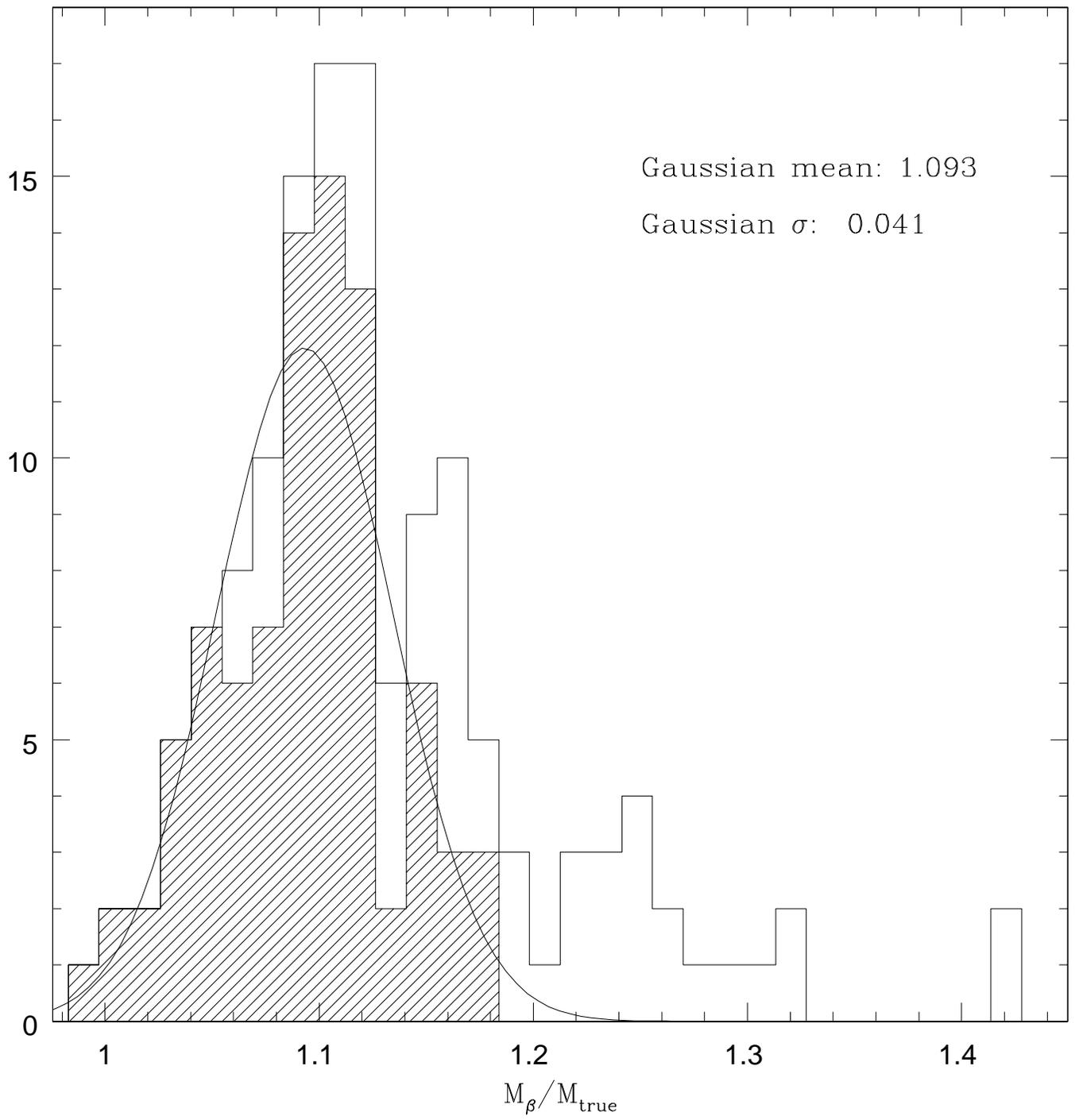
\end{document}